\documentclass[twocolumn,  
showpacs,  
preprintnumbers,  
aps,  
prd,  
a4paper,  
nofootinbib,  
tightenlines,  
floats, floatfix  
]{revtex4}

\usepackage{amsmath}
\usepackage{amsfonts}
\usepackage{graphicx,color}

\newcommand{\nn}{\nonumber \\}
\newcommand{\bea}{\begin{eqnarray}}
\newcommand{\ena}{\end{eqnarray}}

\begin{document}

\preprint{BI-TP 2010/29}

\title{Observational constraints on the energy scale of inflation}

\bigskip

\author{Zong-Kuan Guo}
\email{guozk@physik.uni-bielefeld.de}
\affiliation{Fakult{\"a}t f{\"u}r Physik, Universit{\"a}t Bielefeld, Postfach 100131, 33501
Bielefeld, Germany}

\author{Dominik J. Schwarz}
\email{dschwarz@physik.uni-bielefeld.de}
\affiliation{Fakult{\"a}t f{\"u}r Physik, Universit{\"a}t Bielefeld, Postfach 100131, 33501
Bielefeld, Germany}

\author{Yuan-Zhong Zhang}
\email{zyz@itp.ac.cn}
\affiliation{Institute of Theoretical Physics, Chinese Academy of Sciences, P.O. Box 2735, Beijing 100190, China}

\date{\today}

\begin{abstract}
Determining the energy scale of inflation is crucial to
understand the nature of inflation in the early Universe.
Assuming a power-law power spectrum of primordial curvature perturbations,
we place observational constraints on the energy scale of
the observable part of the inflaton potential
by combining the 7-year Wilkinson Microwave Anisotropy Probe data
with distance measurements from the baryon acoustic oscillations in the distribution of galaxies
and the Hubble constant measurement.
Our analysis provides an upper limit on this energy scale,
$2.3 \times 10^{16}$ GeV at 95\% confidence level.
Moreover, we forecast the sensitivity and constraints
achievable by the Planck experiment by performing Monte Carlo studies on simulated data.
Planck could significantly improve the constraints on the
energy scale of inflation and on the shape of the inflaton potential.
\end{abstract}


\maketitle

\section{Introduction}

Inflation in the early Universe has become the standard model
for the generation of cosmological perturbations in the Universe,
the seeds for large-scale structure and temperature anisotropies
of the Cosmic Microwave Background (CMB).
The simplest scenario of cosmological inflation is based upon
a single, minimally coupled scalar field with a flat potential.
Quantum fluctuations of this inflaton field give rise to a
Gaussian, adiabatic and nearly scale-invariant power spectrum
of curvature perturbations (see Refs.~\cite{lid95,bas05} for reviews).
This prediction is strongly supported by CMB observations from
the Wilkinson Microwave Anisotropy Probe (WMAP)~\cite{kom08,kom10}.

During slow-roll inflation, the potential energy drives an exponential expansion
of the Universe. Detecting this energy scale is crucial to understand how inflation
arises in a fundamental theory of physics. It is known that the amplitude of the
power spectrum of gravitational waves is directly proportional to the energy
scale of inflation~\cite{lid94}. Hence one could use a determination of the
tensor contribution to the temperature and polarization anisotropies of the
CMB to determine this energy scale.

To our knowledge, the first constraint from CMB observations on the
energy scale of inflation has been obtained by Liddle~\cite{lid93}.
More recent constraints from previous releases of the WMAP data on the
Hubble scale or energy scale during inflation have been presented
in~\cite{lea03,sch04,lid06}. If one assumes a specific
inflationary model, in some cases the energy scale can be pinned down,
e.g. for a Landau-Ginzburg potential~\cite{des07}.

The WMAP collaboration has released the results of 7-year
observations~\cite{kom10}. They updated upper limits on
the tensor-to-scalar ratio, but not on the energy scale of inflation.
In this work, we place observational constrains on the
potential energy scale, the first and second derivatives of the
inflaton potential by using the 7-year WMAP data with Gaussian priors
on the Hubble constant $H_0$ and on the distance ratios
of the comoving sound horizon to the angular diameter distances
from the Baryon Acoustic Oscillations (BAO).
Our analysis assumes that inflation is driven by a single, minimally
coupled scalar field and that fluctuations on the observable scales
are generated during an epoch of slow roll of this field (the inflaton).
We obtain upper limits on three potential parameters (Taylor coefficients of
the potential).
It is clear that the WMAP data mainly constrain the amplitude
of tensor modes by the low-$l$ temperature and polarization data.
Compared to WMAP, the Planck satellite is designed to measure
temperature and polarization anisotropies of the CMB to higher
accuracy.
We estimate errors of the potential parameters for the Planck
experiment using the Monte Carlo simulation approach.
As expected, Planck could significantly improve the constraints
on the energy scale of inflation.

This paper is organized as follows. In Section~\ref{sec2},
adopting a Taylor expansion of the inflaton potential, we express the
power spectra and their spectral indices of scalar and tensor modes at the pivot scale
$k_0$ in terms of the values of the potential energy and its first and second
derivatives at $\phi_0$. In Section~\ref{sec3} we obtain observational constraints on the
Taylor coefficients of the inflaton potential from the 7-year WMAP data.
Using a Monte Carlo approach, we analyze the sensitivity of the
Planck experiment w.r.t. these Taylor coefficients in Section~\ref{sec4}.
Section~\ref{sec5} is devoted to conclusions.

\section{The energy scale of inflation}\label{sec2}

In standard slow-roll inflation, a scalar field $\phi$ slowly rolls
down its potential $V(\phi)$.
The condition for inflation requires that the potential energy of
the inflaton field dominates over the kinetic energy. A sufficiently
flat potential for the inflation is required in order to lead to
a sufficient number of e-folds.
On the other hand, when scales relevant to current cosmological observations
cross the Hubble radius during inflation, the change in the value
of the inflaton field is typically small.
Hence the inflationary potential can be expanded as a Taylor series
\bea
\label{vphi}
V(\phi) &=& V_0(\phi_0) + V_1(\phi_0)(\phi-\phi_0) \nn
&& + \frac12 V_2(\phi_0)(\phi-\phi_0)^2 + \cdots
\ena
about the point $\phi_0$. Here $V_0$ is the energy scale of inflation,
$V_1$ is the potential force term that is balanced by the Hubble
friction term during slow-roll inflation, and $V_2$ is the effective
mass term of the inflaton field at $\phi_0$. This approach is different from
the so-called reconstruction of the inflaton potential \cite{lid94},
as we do not track the
evolution of the inflaton; we rather evaluate the potential at the pivot point
$\phi_0$. Let us stress that we have to assume that the Taylor series in Eq.~(\ref{vphi})
converges.

According to the sign of the second derivative of the potential,
inflationary models can be classified into large-field models
and small-field models~\cite{kol99}. The former have $V_2>0$ while
the latter have $V_2<0$ (see Figures~\ref{wmp7}-\ref{r01v}).
Linear models with $V_2=0$ live on the boundary between large-field
and small-field models. However, this classification might be misleading,
as the sign of $V_2$ has nothing to do with the initial conditions of inflation.
The classification is inspired by the Mexican hat potential, which leads to inflation for two types of
initial conditions: false vacuum initial conditions lead to $V_2 <0$, while the initial
conditions of chaotic inflation give $V_2 > 0$. But a counter example is the recently studied
non-minimally coupled Higgs field~\cite{BS}, which in the Einstein frame has an effective potential with
$V_2 < 0$, despite being a large-field model (initial field values beyond the reduced Planck mass).
Nevertheless, the sign of $V_2$ bears physical content. If $V_2 >0$, we can continue the
potential to the point when inflation comes to an end, but if $V_2 < 0$, it is clear that some
transition has to happen in order to finish the quasi-exponetial expansion of the Universe
(see~\cite{sch04}).

We assume that the primordial power spectrum is a power-law.
To leading order in the slow-roll approximation, the power spectra of
the scalar and tensor perturbations, and their spectral
indices can be written in terms of the value of the potential energy
and its first and second derivatives at $\phi_0$
\bea
\label{As}
A_{\rm s} &=& \frac{1}{12\pi^2M^6_{\rm Pl}}\frac{V^3_0}{V^2_1}, \\
r &=& 8M^2_{\rm Pl}\frac{V^2_1}{V^2_0},
\ena
\bea
n_{\rm s}-1 &=& -3M^2_{\rm Pl}\frac{V^2_1}{V^2_0} + 2M^2_{\rm Pl}\frac{V_2}{V_0},\\
n_{\rm t} &=& -M^2_{\rm Pl} \frac{V^2_1}{V^2_0},
\label{nt}
\ena
where $M_{\rm Pl}$ is the reduced Planck mass and $r$ is the
tensor-to-scalar ratio.
The amplitude of the power spectra and spectral indices
are defined at the scale $k_0$, which corresponds to the value
$\phi_0$ of the inflaton field when the mode exits the
horizon during inflation.
Note that only the $V_1^2$ term appears in Eqs.~(\ref{As})-(\ref{nt}), which means that
$\pm V_1$ give the same inflationary variables ($A_{\rm s},r,n_{\rm s},n_{\rm t}$).
Hence observations such as CMB anisotropies cannot determine the
sign of the first derivative of the potential.
This fact can also be understood from the expansion~(\ref{vphi}).
Actually the sign of $V_1$ is unphysical, because changing the sign of $V_1$
is equivalent to $\phi\to-\phi$ and the sign of $\phi$ itself cannot be measured.
In what follows we will consider the case of $V_1>0$.

In terms of the power spectra and the scalar spectral index, the
coefficients in the Taylor expansion~(\ref{vphi}) can be expressed (to leading order)
as (if $r > 0 \Leftrightarrow V_1 \neq 0$)
\bea
\label{v1}
V_0(\phi_0) &=& \frac{3\pi^2}{2} M_{\rm Pl}^4 A_{\rm s} r,\\
V_1(\phi_0) &=& \frac{3\pi^2}{4\sqrt{2}} M_{\rm Pl}^3 A_{\rm s} r^{3/2},\\
V_2(\phi_0) &=& \frac{3\pi^2}{4} M_{\rm Pl}^2 A_{\rm s} r \left[\frac38 r + (n_{\rm s}-1)\right],
\label{v2}
\ena
which can be used to reconstruct the inflationary potential~\cite{lid94}
(see Ref.~\cite{lid95} for a review). In this paper, we use
it to determine the potential parameters when our Universe
crosses the Hubble radius during inflation.
Note that the sign of $V_2$ is determined by the sign of
$(3r/8+n_{\rm s}-1)$, which is consistent with the classification of
inflationary models in the $n_{\rm s}$-$r$ plane~\cite{kol99}.

\section{Observational constraints from WMAP}\label{sec3}

We assume a spatially flat $\Lambda$CDM model as background
cosmology and take the power spectrum parameters ($A_{\rm s},r,n_{\rm s},n_{\rm t}$)
as input parameters.
Then the potential parameters ($V_0, V_1, V_2$) can be derived
by the relation~(\ref{v1})-(\ref{v2}).
The reason why we choose the power spectrum parameters instead
of the potential parameters as input parameters in Markov chains is that
the power spectrum parameters are very sensitive to the values
of the potential parameters especially close to the origin,
which leads to an extremely slow convergence of Markov chains.
Analysis is carried out by using the publicly available CosmoMC
package~\cite{lew02}, which explores the parameter space by
means of Monte Carlo Markov Chains.
Besides the 7-year WMAP data including the low-$l$ temperature
($2 \le l \le 32$) and polarization ($2 \le l \le 23$)
data~\cite{kom10}, we use
two main astrophysical priors: the present-day Hubble constant $H_0$
from the magnitude-redshift relation of 240 low-$z$ Type Ia
supernovae at $z<0.1$~\cite{rie09}, and the angular diameter distances out
to $z=0.2$ and $0.35$, measured from the two-degree field galaxy
redshift survey and the sloan digital sky survey data~\cite{per09}.
Following the WMAP team we choose $k_0=0.002$ Mpc$^{-1}$ as
the pivot scale of the primordial power spectra.

\begin{figure}[!htb]
\begin{center}
\includegraphics[width=8cm]{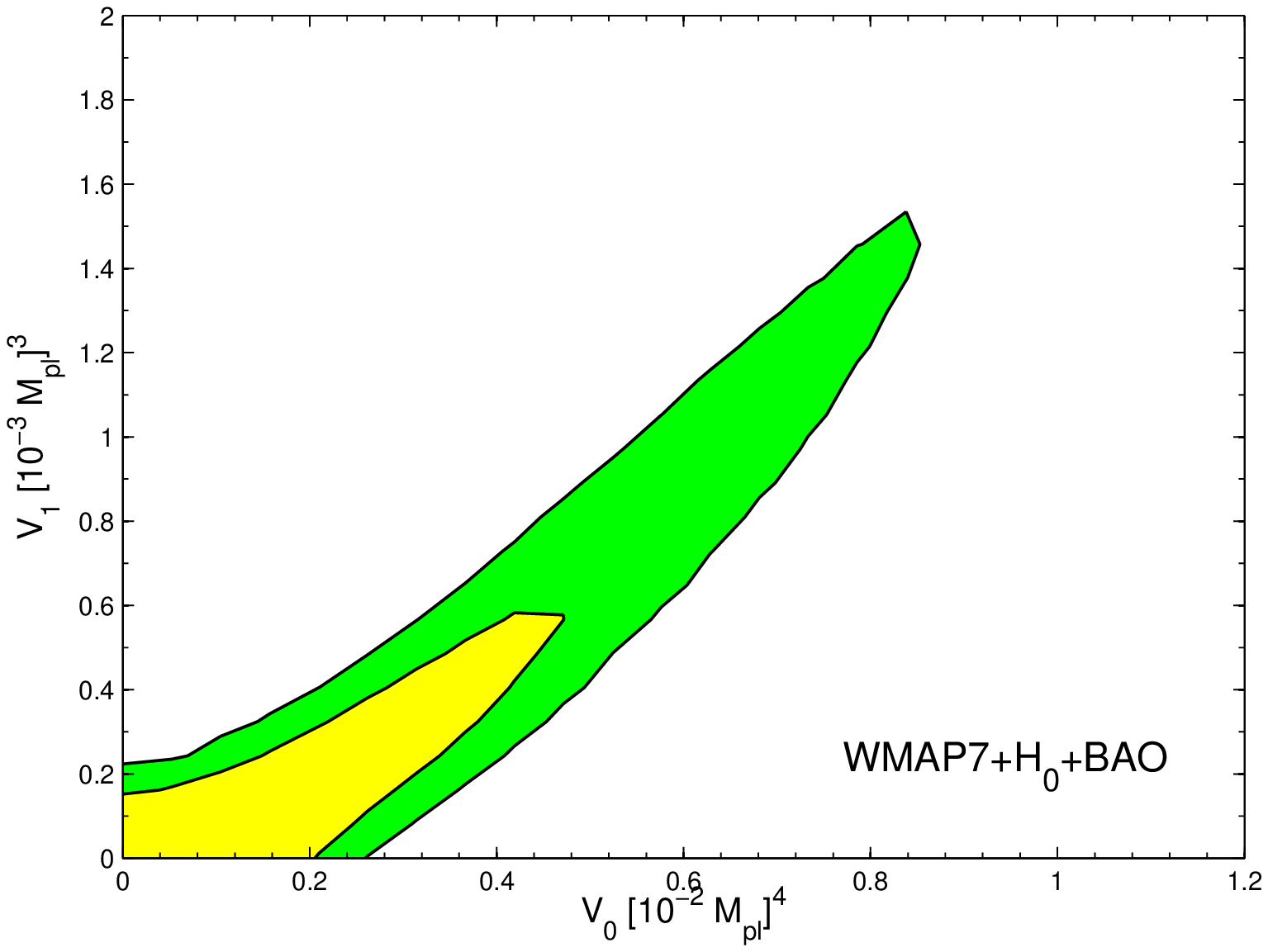}
\includegraphics[width=8cm]{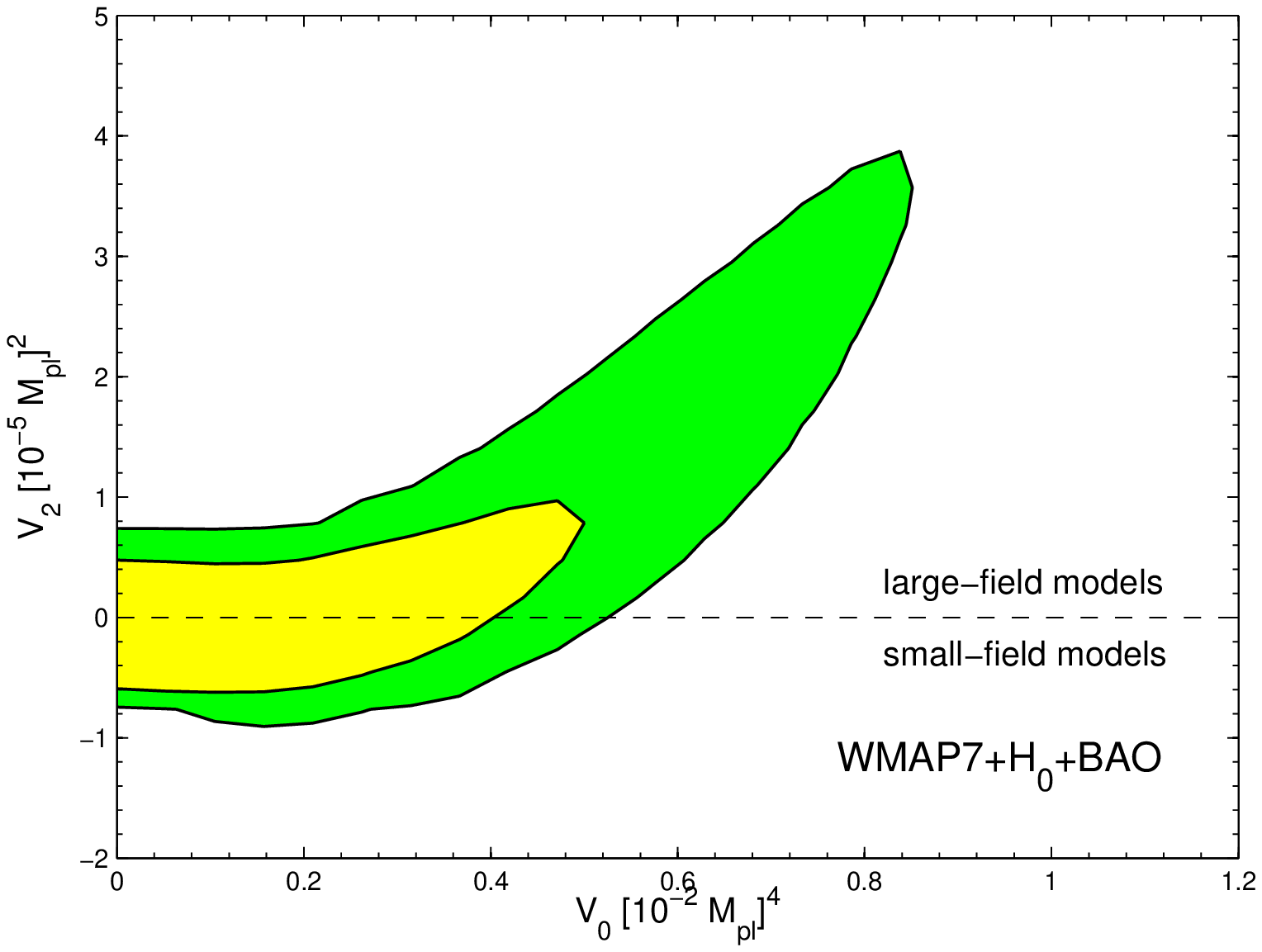}
\includegraphics[width=8cm]{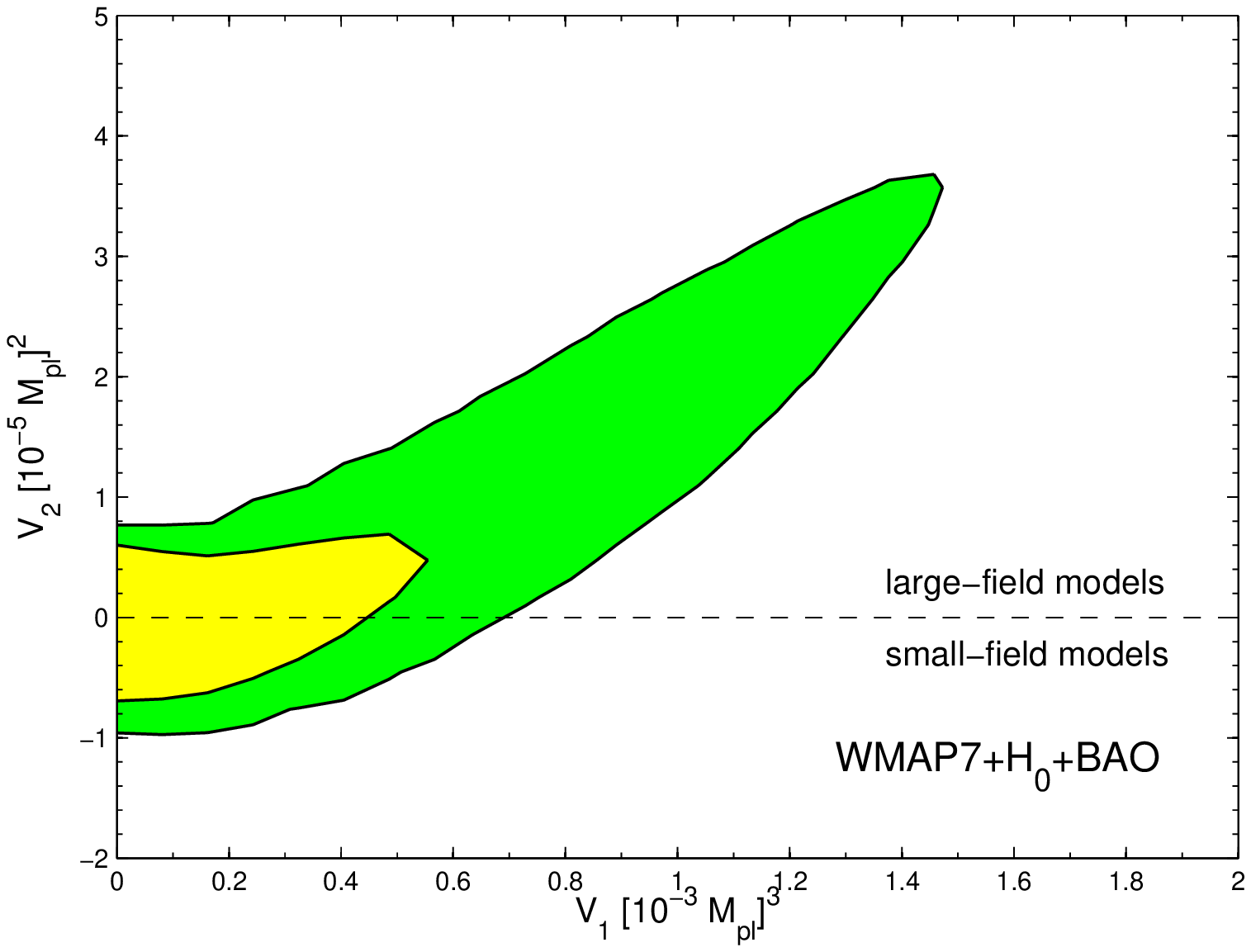}
\caption{Two-dimensional joint marginalized constraints (68\% and
95\% C.L.) on the $V_0$-$V_1$ plane (top panel),
the $V_0$-$V_2$ plane (middle panel) and the
$V_1$-$V_2$ plane (bottom panel) derived from the 7-year WMAP data
with Gaussian priors on $H_0$ and the distance ratios from the BAO.
The dashed line corresponding to linear models marks the border of
large-field and small-field models.}
\label{wmp7}
\end{center}
\end{figure}

Figure~\ref{wmp7} shows the two-dimensional joint marginalized
constraints on $V_0$, $V_1$ and $V_2$ at 68\% and 95\%
confidence level (C.L.) from the 7-year WMAP data with Gaussian priors
on $H_0$ and the distance ratios.
We find upper limits on the potential energy, the first and second
derivative of the potential:
\bea
&& V^{1/4}_0 \lesssim 2.3 \times 10^{16} {\mathrm{GeV}},
 \quad V^{1/3}_1 \lesssim 2.7 \times 10^{15} {\mathrm{GeV}}, \nn
&& |V_2|^{1/2} \lesssim 4.5 \times 10^{13} {\mathrm{GeV}},
\ena
at $95\%$ C.L..
Thus the upper limit on the energy scale of inflation is less than
two orders of magnitude lower than the reduced Planck mass, which
corresponds to the grand unified theory scale. We see that
$V_0^{1/4} > V_1^{1/3} > |V_2|^{1/2}$, which is consistent with our assumption that the
Taylor series converges. Moreover, $|V_2|^{1/2}$ is almost five orders of magnitudes below the
Planck scale, which confirms that the potential must be very flat, consistent with the
assumption of slow-roll inflation. Another observation is that there is a strong degeneracy of
the upper limits of all considered Taylor coefficients on $V_0$. This stresses again
how important the identification of the inflationary energy scale is for model building.
We have also checked that $V_0$, $V_1$ and $V_2$ are not
degenerate with any of the other cosmological parameters.

\section{Future constraints from Planck}\label{sec4}

In this section, following the approach described in
Refs.~\cite{per06,gal10} we generate synthetic data for the Planck
experiment and then perform a systematic analysis on the simulated data.
First of all, we assume a
fiducial cosmological model: baryon density $\Omega_bh^2=0.0227$,
cold dark matter density $\Omega_ch^2=0.108$, the Hubble parameter
$h=72.4$, reionization optical depth $\tau=0.089$, amplitude of
scalar perturbation $A_{\rm s}=2.41\times10^{-9}$, scalar spectral index
$n_{\rm s}=0.961$ and tensor-to-scalar ratio $r=0.1$. Given the fiducial
cosmological model, one can use a Boltzmann code such as
CAMB~\cite{lew99} to calculate the power spectra for the temperature
and polarization anisotropies $C_l^{TT}$, $C_l^{EE}$, $C_l^{TE}$
and $C_l^{BB}$.

We assume that beam uncertainties are small and that uncertainties
due to foreground removal are smaller than statistical errors.
For an experiment with multiple channels $c$ with different beam width
and sensitivity, the noise power spectrum $N_l^{XX}$ can be
approximated as~\cite{kin98}
\bea
\left(N_l^{XX}\right)^{-1} &=& \sum_c \left(\sigma^{(c)}_X \theta^{(c)}_{\mathrm{FWHM}}\right)^{-2} \nn
&& \exp \left[-l(l+1)(\theta^{(c)}_{\mathrm{FWHM}})^2/(8\ln 2)\right],
\ena
where $\sigma^{(c)}_X$ is the root mean square of the instrumental noise
per pixel for temperature ($X=T$), $E$-mode polarization ($X=E$) and
$B$-mode polarization ($X=B$), and
$\theta^{(c)}_{\mathrm{FWHM}}$ is the full width at half maximum of
Gaussian beam for channel $c$.
Non-diagonal noise terms are expected to vanish since the noise
contribution from different maps are uncorrelated.
For Planck we combine only the $100$, $143$ and $217$
GHz HFI channels, with beam width $\theta_{\mathrm{FWHM}}=(9.6',7.0',4.6')$
in arcminutes, temperature noise per pixel $\sigma_T=(8.2,6.0,13.1)$ in $\mu K$
and polarization noise per pixel $\sigma_{E,B}=(13.1,11.2,24.5)$
in $\mu K$ (see Ref.~\cite{tau10} for the instrumental
specifications of Planck).

Given the fiducial spectra $C^{XY}_l$ and noise spectra
$N^{XY}_l=\delta_{XY}N^{XX}_l$,
one can generate a random realization of $a^X_{lm}$ using the following
method~\cite{per06}
\bea
a^T_{lm} &=& \sqrt{\bar{C}^{TT}_l}G^{(1)}_{lm}, \\
a^E_{lm} &=& \frac{\bar{C}^{TE}_l}{\bar{C}^{TT}_l}\sqrt{\bar{C}^{TT}_l}G^{(1)}_{lm}
 +\sqrt{\bar{C}^{EE}_l-\frac{(\bar{C}^{TE}_l)^2}{\bar{C}^{TT}_l}}G^{(2)}_{lm},\\
a^B_{lm} &=& \sqrt{\bar{C}^{BB}_l}G^{(3)}_{lm},
\ena
where $\bar{C}^{XY}_l=C^{XY}_l+N^{XY}_l$ and $G^{(i)}_{lm}$ are
Gaussian-distributed random numbers with unit variance. Then one
can reconstruct the power spectra of the mock data $\hat{C}^{XY}_l$ by
\bea
\hat{C}^{XY}_l = \frac{1}{2l+1} \sum_{m=-l}^l a^{X*}_{lm} a^Y_{lm}.
\ena

Once simulated data are produced we perform a Monte Carlo analysis through
the effective $\chi^2$ defined as~\cite{eas04}
\bea
\chi^2_{\mathrm{eff}} &=& \sum_l (2l+1)f_{\mathrm{sky}} \nn
 && \Bigg\{\frac{\hat{C}_l^{TT}\bar{C}_l^{EE}
 + \bar{C}_l^{TT}\hat{C}_l^{EE}
 -2\hat{C}_l^{TE}\bar{C}_l^{TE}}
 {\bar{C}_l^{TT}\bar{C}_l^{EE} - (\bar{C}_l^{TE})^2}
 +\frac{\hat{C}_l^{BB}}{\bar{C}_l^{BB}} \nn
&& +\ln\frac{\bar{C}_l^{TT}\bar{C}_l^{EE} - (\bar{C}_l^{TE})^2}
 {\hat{C}_l^{TT}\hat{C}_l^{EE} - (\hat{C}_l^{TE})^2}
 +\ln\frac{\bar{C}_l^{BB}}{\hat{C}_l^{BB}}-3 \Bigg\},
\ena
where $f_{\mathrm{sky}}$ is the sky fraction sampled by the experiment
after foregrounds removal. For the Planck data we choose
$f_{\mathrm{sky}}=0.65$, corresponding to a $\pm 20^{\circ}$ Galactic cut.

We consider the $TT$, $TE$ and $EE$ power spectra on scales with
$l \le 2000$.
A measurement of the amplitude of the primordial gravitational
waves would allow a direct determination of the inflationary energy scale.
Both the primordial gravitational waves and the weak gravitational
lensing of the $E$-mode polarization are sources of $B$-mode
polarization~\cite{zal98}.
On large scales, the contributions to the $B$-mode signal mainly
come from the primordial gravitational waves generated by inflation.
The effect of the weak gravitational leansing on the $B$-mode power
spectrum ultimately dominates on small scales.
For the fiducial model with $r=0.1$ the $B$-mode signal generated
by the weak gravitational lensing dominates above $l \sim 150$ and
peaks at $l \sim 1000$.
We consider the $BB$ power spectrum up to the peak of the spectrum
at $l \sim 1000$ since the uncertainty in $C^{BB}_l$ arising from
instrument noise becomes large on smaller scales.

\begin{figure}[!htb]
\begin{center}
\includegraphics[width=8cm]{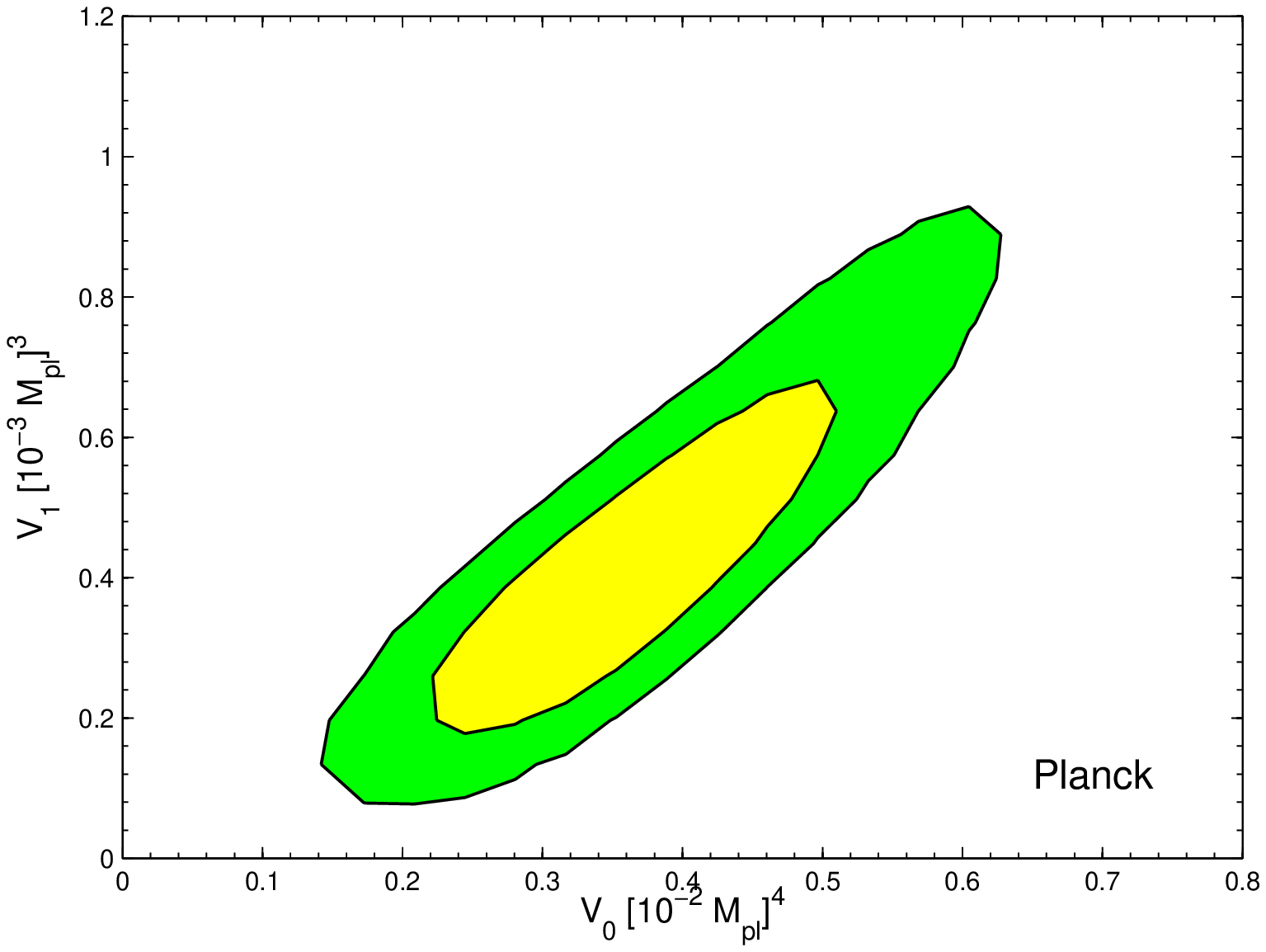}
\includegraphics[width=8cm]{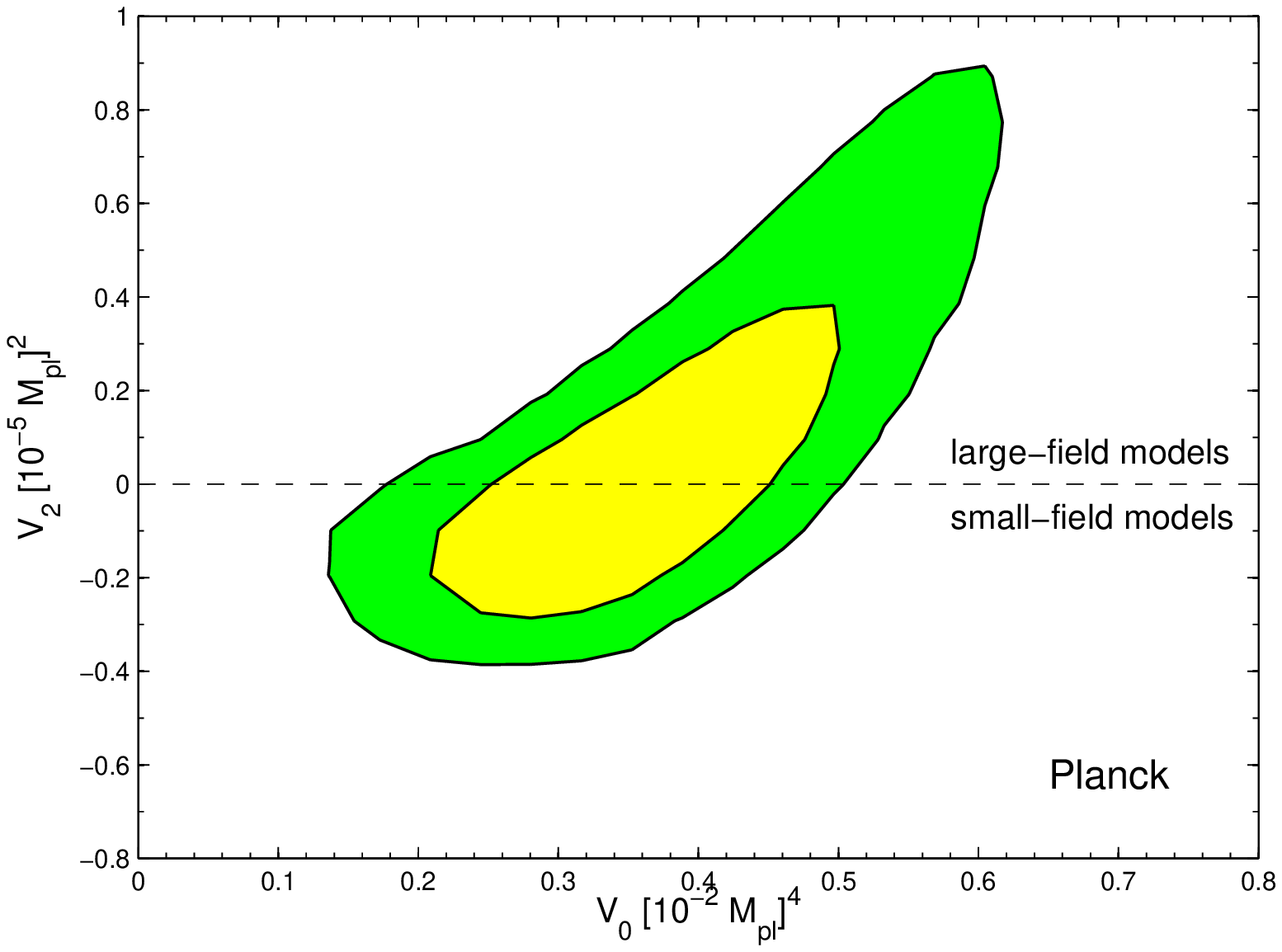}
\includegraphics[width=8cm]{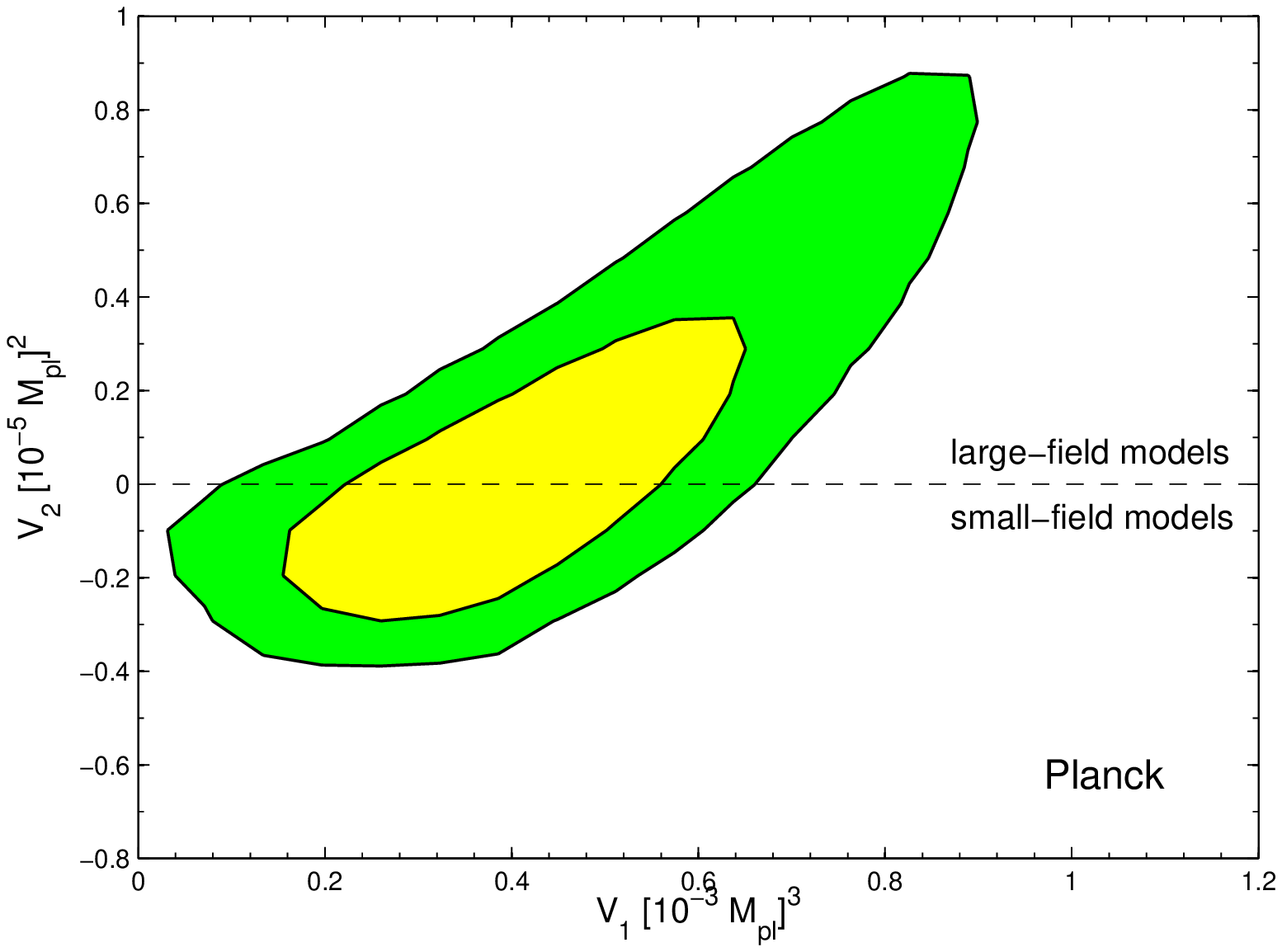}
\caption{Forecast constraints (68\% and 95\% C.L.) on the
$V_0$-$V_1$ plane (top panel), the $V_0$-$V_2$ plane (middle panel)
and the $V_1$-$V_2$ plane (bottom panel)
for the Planck experiment in the case of $r=0.1$.
The dashed line corresponding to linear models marks the border of
large-field and small-field models.}
\label{v0v1v2}
\end{center}
\end{figure}

Figure~\ref{v0v1v2} shows the forecast constraints on $V_0$, $V_1$ and $V_2$
at 68\% and 95\% C.L. from the simulated
Planck data in the case of $r=0.1$.
As expected, it illustrates dramatically how Plank can break
degeneracies between the potential parameters if the
scalar-to-tensor ratio is tightly constrained by Planck.
The marginalized $1\sigma$ (68\%) errors on $V_0$, $V_1$ and $V_2$ are
0.099 $[10^{-2} M_{\rm Pl}]^4$,  0.18 $[10^{-3} M_{\rm Pl}]^3$ and
0.27 $[10^{-5} M_{\rm Pl}]^2$, respectively.
As we can see, Planck can place strong constraints on the inflationary
energy scale, which would provide a firm observational link with
the physics of the early Universe.
The primordial gravitational waves could be detected at $>95\%$ C.L.
based on our fiducial cosmological model with $r=0.1$.
In order to distinguish the large-field models from the
small-field models it is important to detect the sign of the
second derivative of the potential. The analysis of our fiducial cosmological model
indicates that Planck will not be able to distinguish these models.

\begin{figure}[!htb]
\begin{center}
\includegraphics[width=8cm]{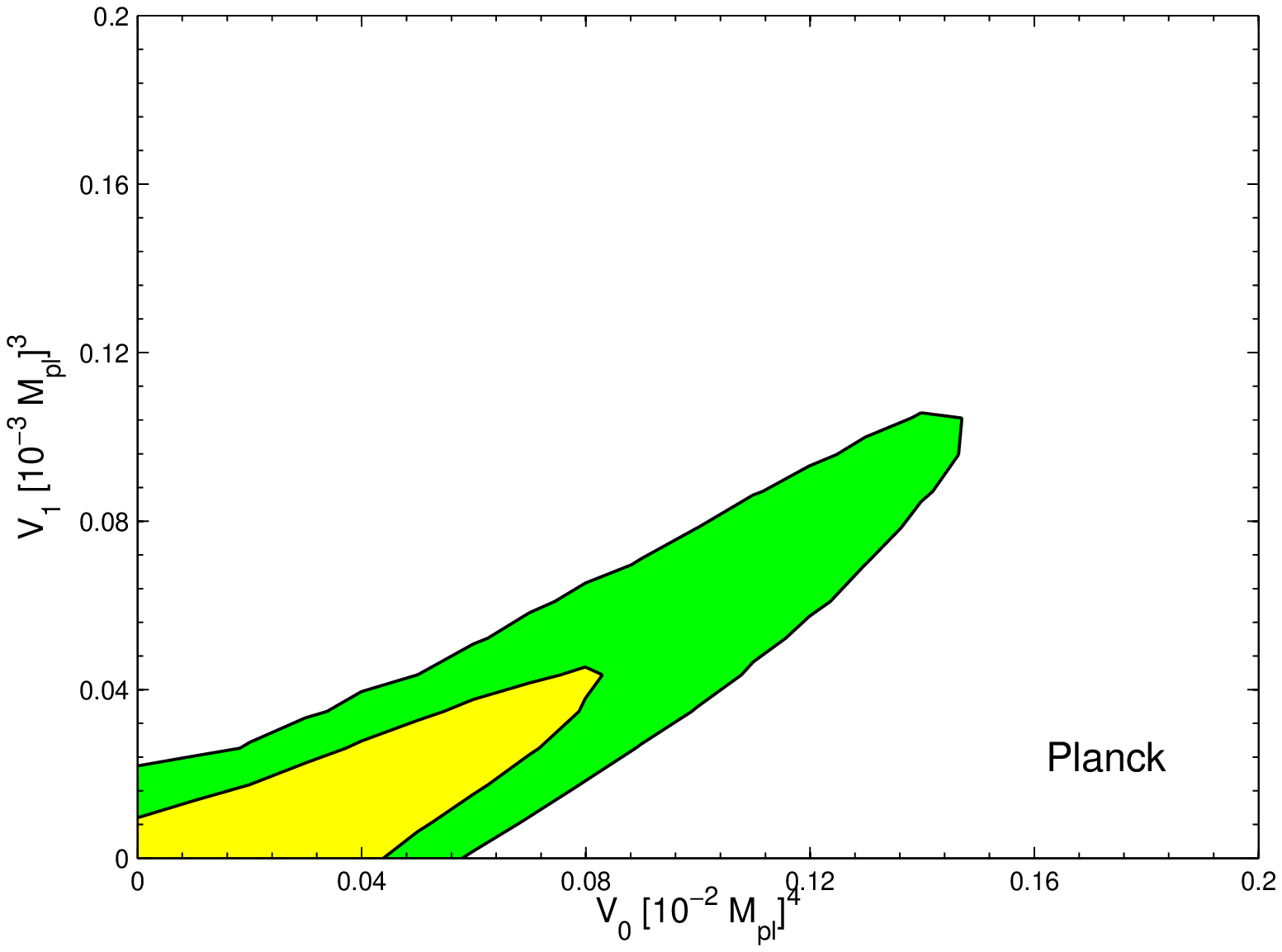}
\includegraphics[width=8cm]{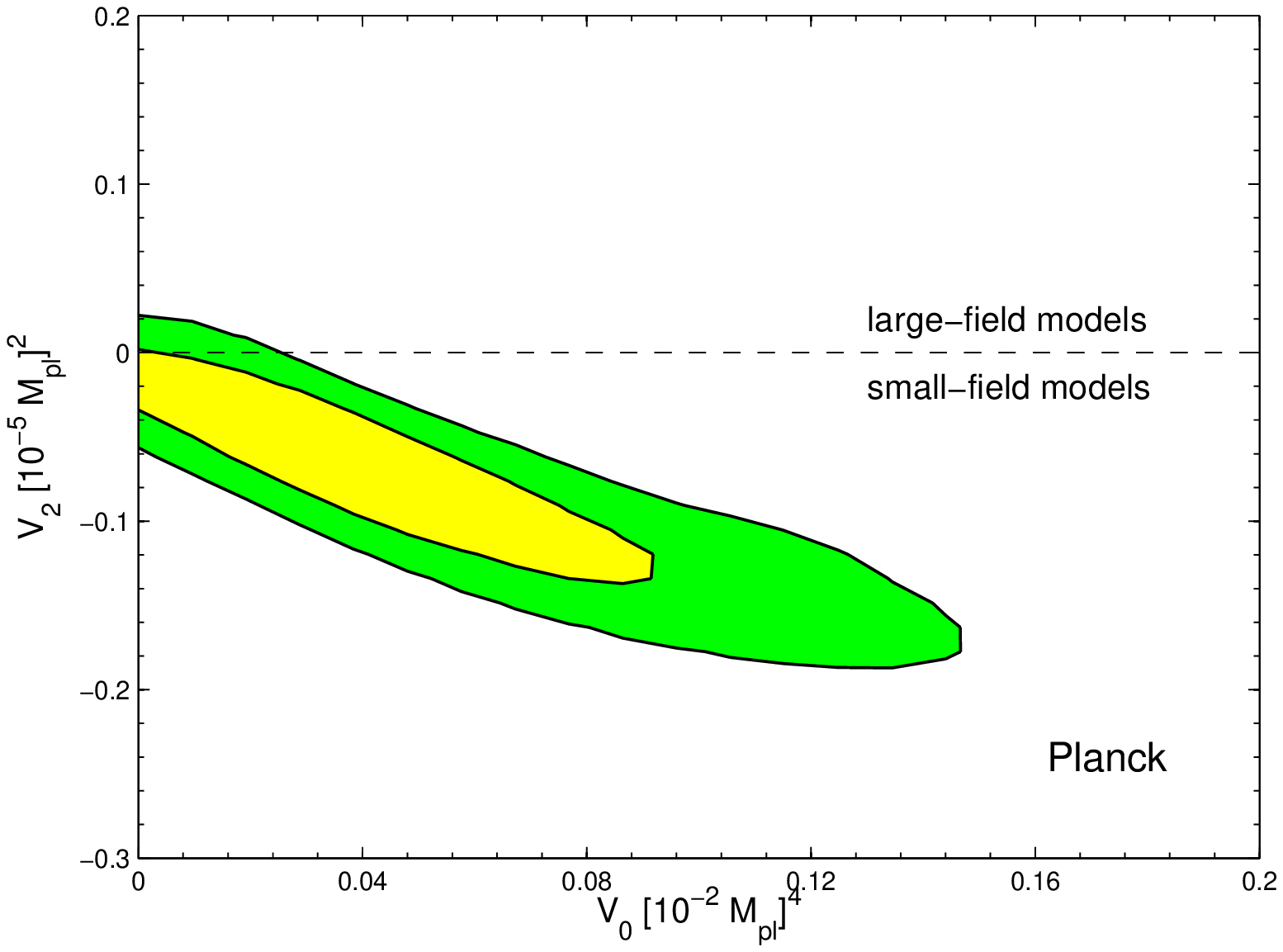}
\includegraphics[width=8cm]{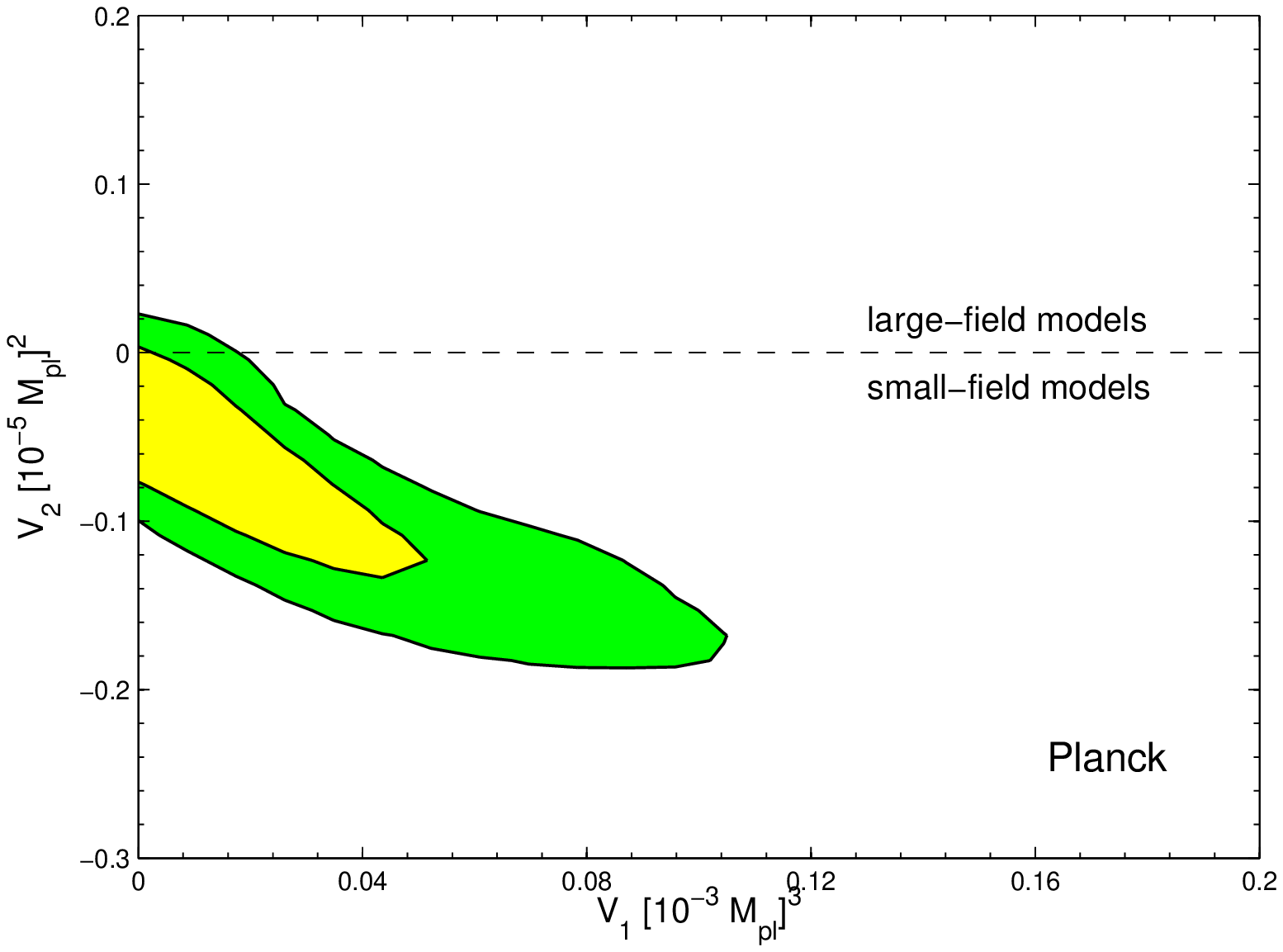}
\caption{Forecast constraints (68\% and 95\% C.L.) on the
$V_0$-$V_1$ plane (top panel), the $V_0$-$V_2$ plane (middle panel)
and the $V_1$-$V_2$ plane (bottom panel)
for the Planck experiment in the case of $r=0.01$.
The dashed line corresponding to linear models marks the border of
large-field and small-field models.}
\label{r01v}
\end{center}
\end{figure}

What about the forecast constraints from Planck if $r\ll 0.1$?
Here we show the case of $r=0.01$.
The fiducial values of the other parameters are
taken to be the WMAP5 maximum likelihood values~\cite{kom08}.
Figure~\ref{r01v} shows the forecast constraints on the potential parameters
at 68\% and 95\% C.L. from the simulated
Planck data in the case of $r=0.01$.
We can see that in this case Planck only provides the upper limits on the
energy scale of inflation as WMAP7 does.
It is implied that Planck fails to detect the primordial gravitational waves if the
tensor-to-scalar ratio is less than one percent.
As shown in Figure~\ref{r01v}, in this case ``small-field" models would obviously be
favored by data compared to the case of $r=0.1$.
It can be understood from Eq.~(\ref{v2}).
If $r \sim 0.01$ and $n_s \sim 0.96$, the second term dominates
over the first one in the bracket. Therefore, a red tilt of the power
spectrum of curvature perturbations leads to a negative
second derivative of the potential.
The upper limits achievable by Planck are
\bea
&& V_0^{1/4} \lesssim 1.4 \times 10^{16}{\rm GeV}, \quad
V_1^{1/3} \lesssim 1.1 \times 10^{15}{\rm GeV}, \nn
&& |V_2|^{1/2} \lesssim 9.8 \times 10^{12}{\rm GeV},
\ena
at 95\% C.L..

\section{Conclusions}\label{sec5}

In this paper we have placed observational constrains on the
potential energy scale, the first and second derivative of the
potential by using the 7-year WMAP data, combined with the
latest distance measurements from the baryon acoustic
oscillations in the distribution of galaxies and measurement
of the present-day Hubble constant from supernova data.
A previous upper limit from the first WMAP data release,
combined with large scale structure data from the 2dF galaxy redshift
survey found $V_0^{1/4} \lesssim 2.7 \times 10^{16}$~GeV at 90\% C.L.~\cite{lea03,sch04}.
Our new upper limit on the energy scale of inflation is only slightly stronger
$V_0^{1/4} \lesssim 2.3 \times 10^{16}$~GeV at 95\% C.L., and shows a
degeneracy with the upper limit on the first derivative of the inflaton potential,
$V_1^{1/3} \lesssim 2.7 \times 10^{15}$ GeV at 95\% C.L..
Adding information on the power spectrum of large scale structure,
the limit would be improved slightly~\cite{fin10}.
Both upper limits lie at the
energy scale of grand unified theories and are consistent with scenarios in
which inflation starts very close to the Planck scale. However, a scenario in which there is
just enough inflation~\cite{RS} in order to solve the horizon and flatness problems but
$V \ll M_{\rm Pl}$ for all field values cannot be excluded, nor can false vacuum initial
conditions be excluded. Latter scenarios can be constrained by a lower limit on the reheating
temperature of the Universe. A recent analysis~\cite{MR} of the 7-year WMAP data finds
that the so-called reheating parameter, fully characterizing
the inflationary reheating epoch, is constrained by data,
which yields lower bounds on the reheating temperature, typically
$> 10^2 \rm{GeV} - 10^2 \rm{TeV}$, dependent on inflation scenarios.

The Planck experiment will soon provide a very accurate
measurements of CMB temperature and polarization anisotropies.
Using the Monte Carlo simulation approach, we have presented
forecasts for improved constrains from Planck.
Our results indicate that the degeneracies between the potential
parameters are broken because of the improved constraint on the
tensor-to-scalar ratio from Planck.
Besides Planck also EBEX, a balloon-borne CMB polarization experiment,
has good chances to significantly improve our understanding
of inflation~\cite{rei07}.

In our analysis, we adopt the three-parameter parametrization
of the potential for the case of a single, minimally coupled inflaton field.
To leading order in the slow-roll approximation, it is
consistent to expand the potential only to quadratic order,
because the third derivative corresponds to higher-order
slow-roll parameters.
The advantage of the model-independent parametrization is that
we could discriminate inflationary models by detecting the sign
of the second derivative of the potential.
If we drop one of the other assumptions,
e.g. by means of a coupling of the inflaton and a Gauss-Bonnet term, we change the
prediction of the tensor-to-scalar ratio~\cite{GS} and thus the limit on the inflationary energy scale.

It is quite surprising that one of the most fundamental questions in the context of inflationary cosmology,
namely what is the energy scale of the observable part of the inflationary expansion, has received
relatively little attention in the attempt to extract information from cosmic data. We hope that
we can convince the reader, that even without a detection of tensor fluctuations, some interesting
upper limits on three parameters can be extracted and that our analysis turns out to be fully self-consistent.

\section*{Acknowledgments}
We thank H.-T. Ding and L. Perotto for useful discussions.
Our numerical analysis was performed on the HPC cluster of the RWTH Aachen.
This work was supported in part by the Alexander von Humboldt Foundation.
YZZ is partially supported by National Basic Research Program of China under Grant No:2010CB832805.
We used CosmoMC and CAMB.
We also acknowledge the use of WMAP data from LAMBDA server.

\end{document}